\documentclass[12pt]{article} 
\pdfoutput=1
\usepackage{amssymb,graphicx}
\usepackage{epstopdf}
\usepackage{amsmath,amsfonts}
\usepackage{epsfig} 
\usepackage{graphicx,graphics}
\usepackage{xcolor}

\begin{document} 

\sloppy

\title{\bf Gravitational misalignment mechanism of Dark Matter production}
\author{Eugeny Babichev$^{a}$, Dmitry Gorbunov$^{b,c}$, and Sabir Ramazanov$^{d}$\\
 \small{$^a$\em Universit\'e Paris-Saclay, CNRS/IN2P3, IJCLab, 91405 Orsay, France }\\
  \small{$^b$\em Institute for Nuclear Research of the Russian Academy of Sciences,}\\
\small{\em 60th October Anniversary prospect 7a, Moscow 117312, Russia}\\
\small{$^c$ \em Moscow Institute of Physics and Technology, Institutsky per. 9, Dolgoprudny 141700, Russia}\\
\small{$^d$ \em CEICO, Institute of Physics of the Czech Academy of Sciences,}\\
\small{\em Na Slovance 1999/2, 182 21 Prague 8, Czech Republic}
 }

{\let\newpage\relax\maketitle}

\begin{abstract}
We consider Dark Matter composed of an oscillating singlet scalar field. On top of the mass term, the scalar is equipped with a potential spontaneously breaking $Z_2$-symmetry. This potential dominates 
at early times and leads to the time-dependent expectation value of the scalar, which decreases in the expanding Universe. As it drops below some critical value, the symmetry gets restored, and the Dark Matter field starts to oscillate around zero.  
We arrange the spontaneous symmetry breaking through the interaction of the scalar with the Ricci curvature. In that way, superheavy Dark Matter can be produced at very early times. Depending on its mass, 
the production takes place at inflation (very large masses up to the Grand Unification scale), at preheating, or at radiation-dominated stage (masses $10^{6}-10^{7}~\mbox{GeV}$).

\end{abstract}

\section{Introduction}

Dark Matter (DM) is manifested only through its gravitational interactions. Therefore, it is natural to assume that DM was produced via a mechanism involving gravity only. We introduce a singlet scalar field $\chi$ to
play the role of DM. The DM field $\chi$ is initially in the spontaneously broken phase. We organize 
symmetry breaking through the interaction of the DM field $\chi$ with a slowly changing function $F(x^\mu)$, which is approximately 
homogeneous in the early Universe, $F(x^\mu) \approx F(t)$,
\begin{equation}
\nonumber 
V_{sb} \propto \left(\chi^2-F(t) \right)^2 \; .
\end{equation}
Besides this symmetry breaking potential $V_{sb}$, the field $\chi$ has a standard mass term. As the function $F(t)$ decreases, the mass term becomes more and more relevant. 
At some point, the symmetry gets restored, and the DM field $\chi$ starts oscillating around zero. 
Therefore, this scenario is a type of the misalignment mechanism, where the DM field is originally offset due to the non-trivial function $F(t)$. 
This DM is stable because of the imposed $Z_2$-symmetry. 
As only gravitational interactions of DM are known so far, it is natural to expect that the function $F(t)$ is of gravitational origin.

In the present paper we discuss the case with $F \propto R$, where $R$ is the Ricci scalar. In this model we show that DM with the right abundance can be produced, provided that 
the field $\chi$ is superheavy. 
In particular, DM with the masses up to $10^{16}~\mbox{GeV}$, i.e., of the order of the Grand Unification scale, can be produced at inflation. 
The proposed mechanism of superheavy DM generation differs from previously known ones at least in two aspects. First, generation takes 
place ($16-20$ e-folds) {\it before} the end of inflation. This is to be compared with DM generation through the minimal coupling to gravity~\cite{Chung:1998ua, Chung:1998zb, Kuzmin:1998uv, Kuzmin:1998kk}, which 
occurs at the transition to post-inflationary stage, or DM production at (p)reheating~\cite{Chung:1998ua, Greene:1997ge, Chung:1998rq, Garny:2015sjg}. Second, for the mechanism discussed in this paper, DM production is independent 
of particularities during post-inflationary evolution, such as the reheating temperature of the Universe, efficiency of parametric resonance, rate of inflaton change at the end of inflation~(cf., Ref.~\cite{Babichev:2018mtd}) etc. 
Note that applicability 
of our mechanism is not limited to inflation, but it can also operate at preheating or radiation-dominated stage. In the latter case, DM masses are in the 
range $10^{6}-10^{7}~\mbox{GeV}$, which is of interest from the viewpoint of high-energy IceCube neutrino observations~\cite{Aartsen:2018mxl}.

The outline of the paper is as follows. In Section~2, we discuss generic features of the class of models of interest, 
considering  arbitrary function $F(t)$. In Section~3, we specify the function $F(t)$ assuming that it is 
proportional to the Ricci scalar $R$. There we outline a region in the model parameter space leading to the right 
abundance of DM comprised of the field $\chi$. We show that the mechanism works for superheavy DM with  masses 
$10^{6}~\mbox{GeV} \lesssim M \lesssim 10^{16}~\mbox{GeV}$, and
production takes place at very early times.
We end up with phenomenological prospects of the model in Section~4. 

\section{Generalities}

We start with the following generic Lagrangian, which describes dynamics of the DM field $\chi$:
\begin{equation}
\label{modelbasic}
{\cal L}=\frac{ (\partial_{\mu} \chi)^2}{2} -\frac{M^2 \chi^2}{2}  -\frac{\lambda}{4} \cdot \left[\chi^2 -F(x^\mu) \right]^2 \; .
\end{equation}
Here $F(x^\mu)$ is a positive definite slowly decreasing function, at least at early times.
As we are interested primarily in the background evolution of the field $\chi$, 
one can treat $F(x^\mu)$ as a function of time only, $F(x^\mu)=F(t)$. 
Hereafter, we assume that the function $F(t)$ varies only due to the cosmic expansion, 
\begin{equation}
\label{Frate}
|\dot{F}| = \kappa H  F \; ,
\end{equation} 
where $H$ is the Hubble rate, and the dimensionless quantity $\kappa \lesssim 1$ is a function of the equation of state of the dominant matter in the Universe. We assume that the field $\chi$ has no direct interactions with
components of the Standard Model of particle physics. Given also $Z_2$-symmetry of the Lagrangian~\eqref{modelbasic}, this guarantees the DM stability. 

\begin{figure}[tb!]
\begin{center}
\includegraphics[width=0.8\columnwidth,angle=0]{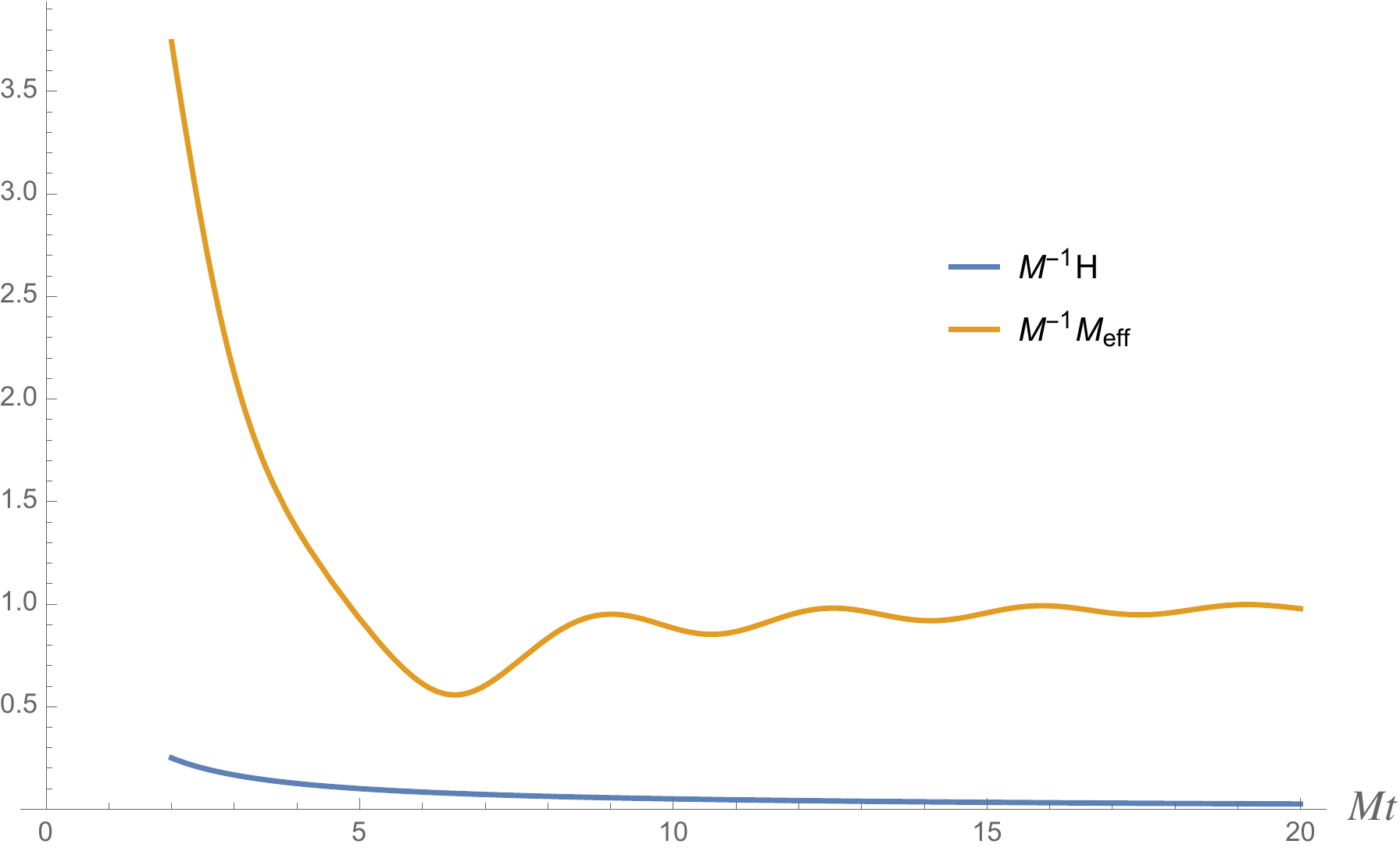}
\caption{The effective mass $M_{eff}$ of the field $\chi$ (orange) and the Hubble parameter (blue) are shown for the model described by Eqs.~(\ref{modelbasic}),~\eqref{Ftoy}, and~\eqref{example}.
Initially, $M_{eff}$ is large and slowly decreases, as the Universe is expanding. 
In this regime the adiabaticity condition~\eqref{adiabaticity} is valid. 
As $M_{eff}$ drops below the bare mass $M$, the adiabaticity condition is violated. From this time on, the field $\chi$ oscillates and it has the constant mass, $M_{eff} \rightarrow M$. 
The effective mass $M_{eff}$ always remains larger than the Hubble parameter. The choice of parameters $\lambda=1/4$ and $\beta=1000$ is assumed.}\label{effectivemass}
\end{center}
\end{figure}

At early times $t$, when $\lambda F(t)>M^2$,
$Z_2$-symmetry is spontaneously broken. Without loss of generality 
we choose the minimum as 
\begin{equation}
\label{minimumbefore}
\chi_{min} (t) =\sqrt{ F(t)-\frac{M^2}{\lambda}} \; .
\end{equation}
For the field $\chi$ to track the minimum, one requires that the latter varies slowly. 
To quantify this statement, consider the effective mass squared of the field at the minimum $\chi=\chi_{min}$ in the spontaneously broken phase:
\begin{equation}
\label{massbreaking} 
M^2_{eff} =  2\left(\lambda F(t)-M^2\right) \; .
\end{equation} 
Let the field start evolution at the values not far from the minimum $\chi_{min}$. This initial relaxation to $\chi_{min}$ happens automatically during inflation, as we will see in what follows. Then, if the effective mass $M_{eff}$ changes slowly with time, 
\begin{equation}
\label{adiabaticity}
\frac{|\dot{M}_{eff} (t)|}{M^2_{eff} (t)} \ll  1 \; ,
\end{equation}
the field $\chi$ resides in the minimum, $\chi \approx
\chi_{min}$. As it follows from the condition~\eqref{adiabaticity}, at the first
stage of evolution, when $M^2_{eff} \approx 2\lambda F(t)$, the effective mass $M_{eff}$ should be much larger than the Hubble rate (see Fig.~\ref{effectivemass}):
\begin{equation}
\label{noisocurvature}
M_{eff} \gg H \; .
\end{equation}
The latter condition is also necessary for the following reason. It guarantees that DM isocurvature perturbations are suppressed during inflation, in agreement with CMB observations~\cite{Akrami:2018odb}. 
Namely, for a large effective mass (relative to the Hubble rate) of the field $\chi$, isocurvature perturbations decay fast in the inflationary Universe.

The l.h.s. of the inequality~(\ref{adiabaticity}) eventually becomes singular when the symmetry breaking minimum~(\ref{minimumbefore}) ceases to exist. Therefore, at some moment of time $t=t_*$ the condition~(\ref{adiabaticity}) is violated
\begin{equation}
\label{adiabaticityviolation}
\frac{|\dot{M}_{eff,*} |}{M^2_{eff,*}} \simeq 1 \; ;
\end{equation}
the subscript $'*'$ refers to the moment of time $t_*$. For times greater than $t_*$ the symmetry is restored, 
and the new minimum is located at $\chi=0$. As the scalar field is offset from the new minimum at $t=t_*$, it oscillates around $\chi=0$ at later times, provided that the bare mass is much larger than the Hubble parameter at $t=t_*$,
\begin{equation}
\label{MHcondition}
M\gg H_* \; .
\end{equation}
We will assume the condition~(\ref{MHcondition}) from now on. The amplitude of the oscillations reads
\begin{equation}
\nonumber 
\chi (t) =\chi_{*} \left(\frac{a_*}{a(t)} \right)^{3/2} \; ,
\end{equation}  
where $a(t)$ is the scale factor; the subscript $'*'$ refers to the moment of time $t_*$. Let us estimate the amplitude $\chi_*$ at the onset of oscillations. From Eqs.~\eqref{massbreaking} and~\eqref{adiabaticityviolation} we obtain
\begin{equation}
\label{effmonset}
M^3_{eff,*}  \simeq \kappa \lambda  F_* H_* \; .
\end{equation}
In particular, this means that the inequality~\eqref{noisocurvature} holds down to the moment $t_*$, when oscillations start. 

From Eqs.~(\ref{massbreaking}) and (\ref{adiabaticityviolation}) it follows that 
\begin{equation}
\label{osconset}
\lambda F_* \simeq M^2 \; .
\end{equation}
Crucially, this equality is not exact: transition to the oscillatory regime occurs, when $\lambda F_ *$ is slightly larger than $M^2$. 
The combination of Eqs.~\eqref{minimumbefore},~\eqref{effmonset}, and~\eqref{osconset} yields for the amplitude $\chi_*$ at the onset of oscillations 
\begin{equation}
\label{chistar}
\chi_* \simeq \frac{\left(\kappa M^{2} H_{*}\right)^{1/3}}{\sqrt{2\lambda}} \; .
\end{equation}
In fact, the same estimate up to factor two can be obtained from the comparison of the kinetic and potential terms in the equation of motion of the field $\chi$. 
Indeed, the field $\chi$ starts to oscillate at the time $t_*$, when $|\ddot{\chi}| \sim \lambda \chi^3_*$. 
Then, using Eq.~\eqref{minimumbefore} one arrives at the estimate~\eqref{chistar}. 

\begin{figure}[tb!]
\begin{center}
\includegraphics[width=0.8\columnwidth,angle=0]{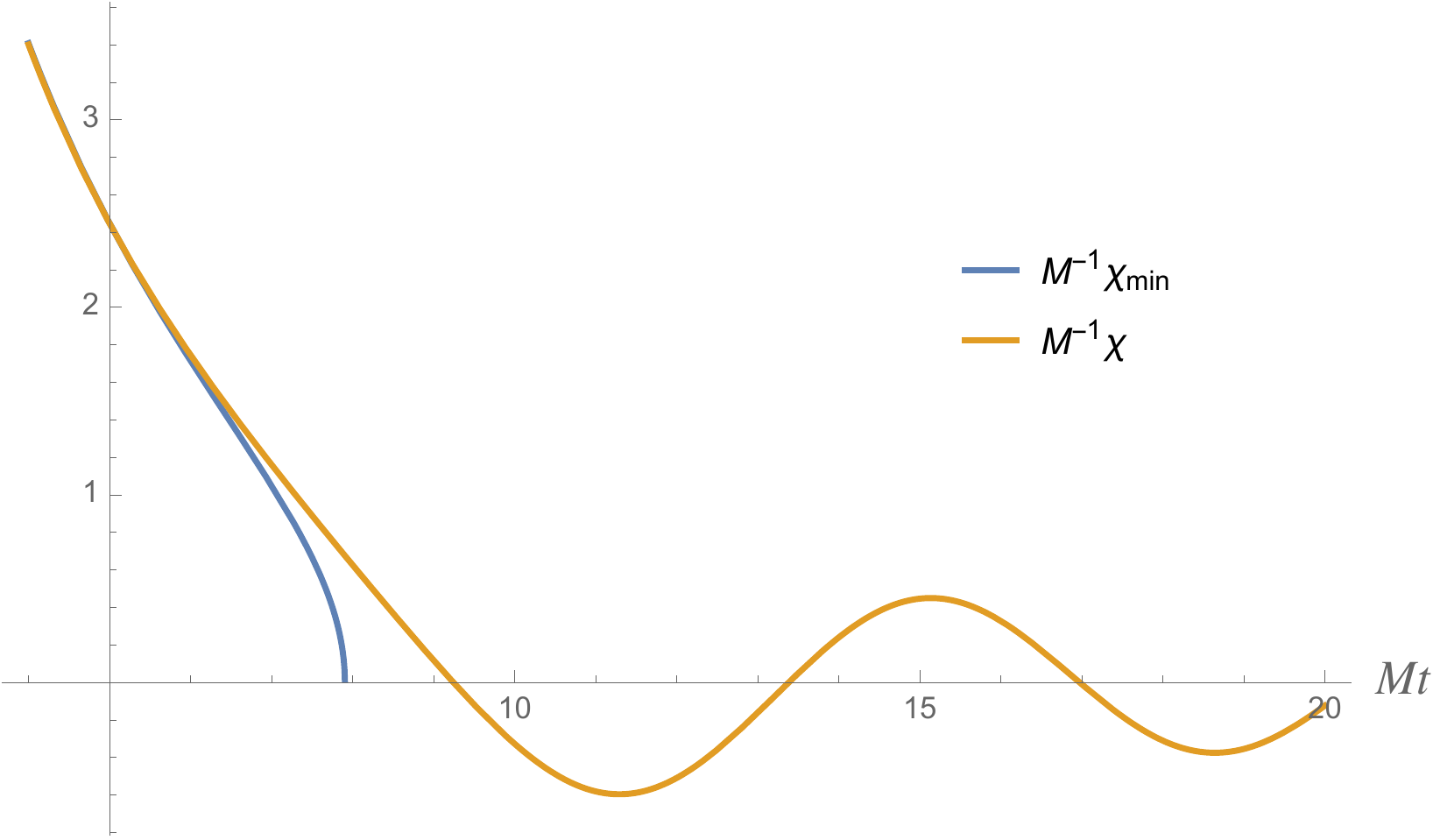}
\caption{Evolution of the DM field $\chi$ (orange)  and the minimum $\chi_{min}$ (blue) are shown for the model described by Eqs.~(\ref{modelbasic}),~\eqref{Ftoy}, and~\eqref{example}. 
At early times, the field $\chi$ resides in the minimum, $\chi=\chi_{min}$, which monotonously decreases with time following the Hubble drag. 
At $t\sim t_*$ the field $\chi$ gets offset from its minimum, and the DM field $\chi$ starts oscillating. The choice of parameters $\lambda=1/4$ and $\beta=1000$ is assumed.}\label{decoupling}
\end{center}
\end{figure}

\begin{figure}[tb!]
\begin{center}
\includegraphics[width=0.45\columnwidth,angle=0]{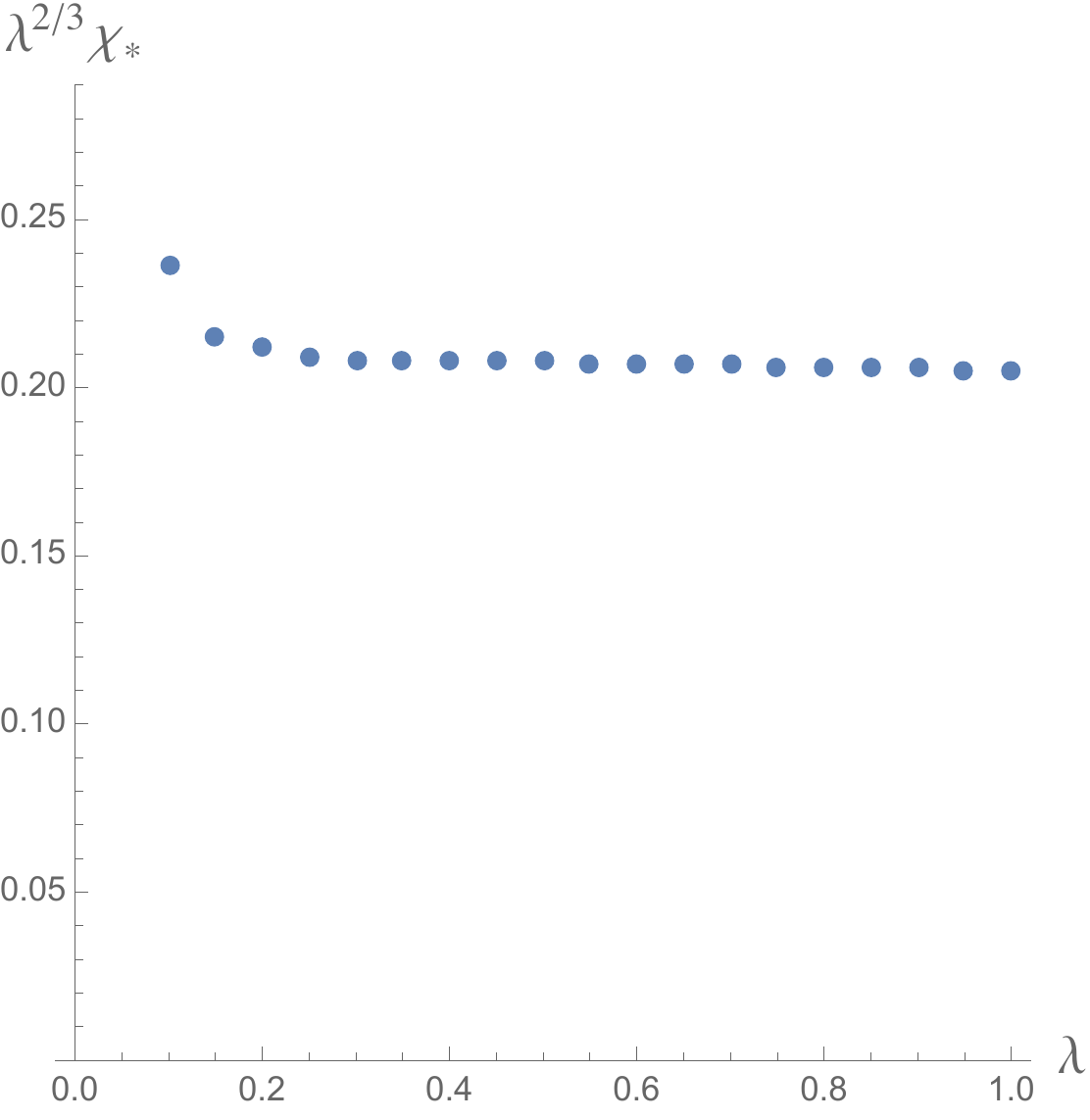}
\includegraphics[width=0.45\columnwidth,angle=0]{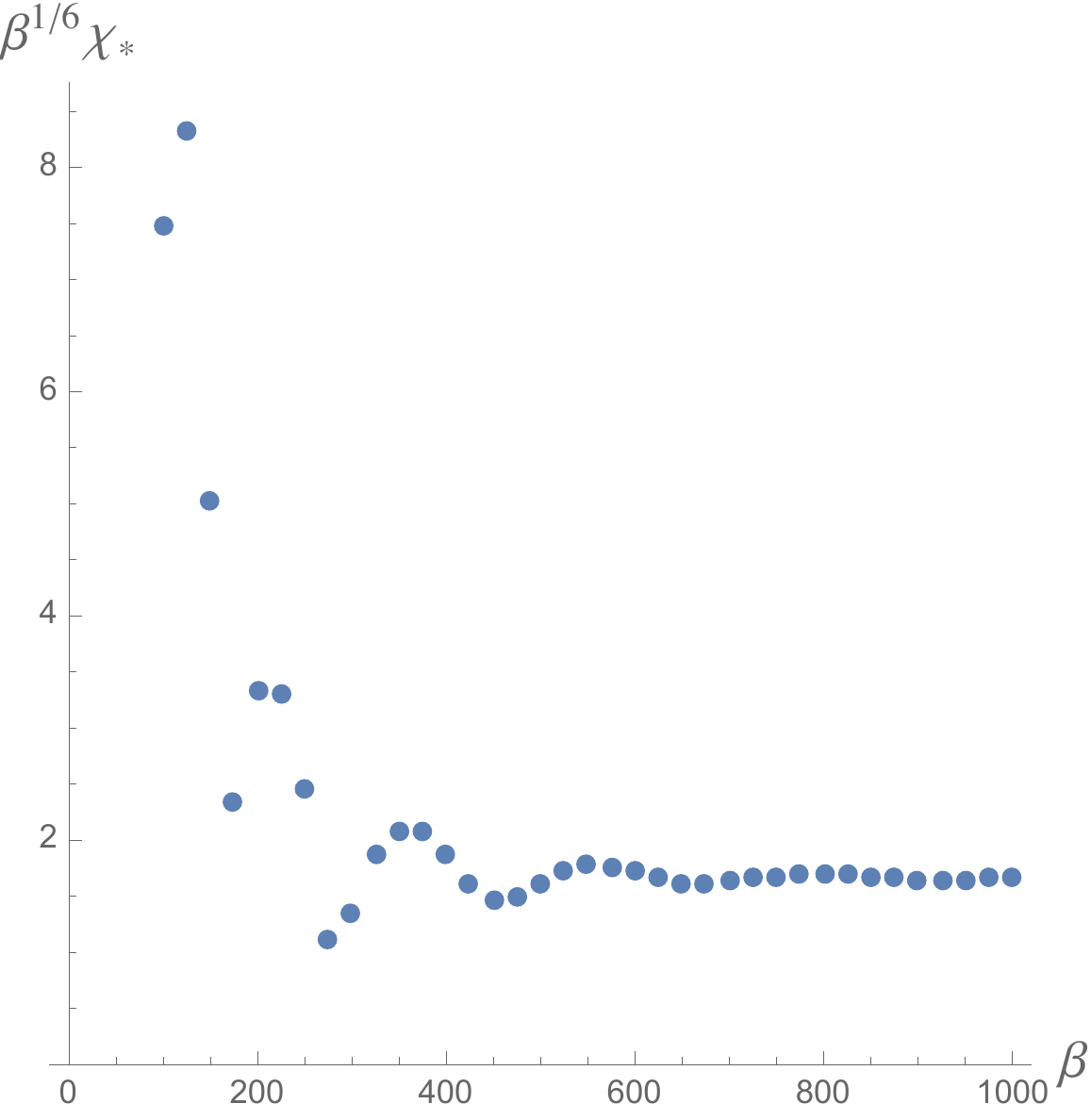}
\caption{The DM field $\chi_*$ at the onset of oscillations is plotted as a function of the parameters of the toy model described by Eqs.~(\ref{modelbasic}),~\eqref{Ftoy}, and~\eqref{example}. 
Dependence on the parameter $\lambda$ is shown for $\beta=1000$ on the left panel; while the right panel shows dependence on the parameter $\beta$ for $\lambda=1/4$. 
For both panels, the behavior of $\chi_*$ matches the analytic expression~\eqref{chistar} for relatively large $\lambda$ and $\beta$, 
when the adiabaticity condition is fulfilled at early times.}\label{lambda}
\end{center}
\end{figure}

Below we illustrate the behavior of the DM field $\chi$ assuming the function $F(t)$ of the form
\begin{equation}
\label{Ftoy}
F(t)=\beta H^2 (t) \; ,
\end{equation}
where $\beta$ is a dimensionless constant. As for the Hubble rate $H(t)$, we choose the following toy example:
\begin{equation}
\label{example}
H=\sigma \left(1-\tanh (\sigma t) \right)+\frac{1}{2\sqrt{t^2+\frac{1}{\sigma^2}}} \; ,
\end{equation}
where $\sigma$ is a dimensionful constant. At the times $t \lesssim -\sigma^{-1}$ (note that $t$ can be both negative and positive), the Hubble rate is nearly constant and 
it models the quasi-de Sitter expansion of the Universe, i.e., inflation. At 
$t \gtrsim \sigma^{-1}$, the Hubble rate $H \simeq \frac{1}{2t}$ describes the subseqeunt stage of radiation domination. In Figs.~\ref{effectivemass} and~\ref{decoupling}, we show evolution of the effective mass 
$M_{eff}$ and the field 
$\chi$ in the toy model described by Eqs.~(\ref{modelbasic}),~\eqref{Ftoy},~and~\eqref{example}. In Fig.~\ref{decoupling}, one can see that the field $\chi$ indeed tracks the minimum $\chi_{min} (t)$ until the moment $t_*$, 
when the adiabaticity condition gets violated. Then, $\chi$ starts to oscillate. 
In Fig.~\ref{lambda}, we plot numerical results for dependence of the field $\chi$ at the onset of oscillations on the parameters $\lambda$ and $\beta$. 
The analytic estimate from Eqs.~\eqref{osconset},~\eqref{chistar}, and~\eqref{Ftoy} gives $\chi_* \sim \lambda^{-2/3} \cdot \beta^{-1/6}$. This is indeed the behavior, which we observe in Fig.~\ref{lambda}. 
Note that the effective mass of the scalar field always remains much larger than the Hubble parameter. 

Consequently, the energy density of DM comprised of the field $\chi$ is estimated as
\begin{equation}
\label{dmenergygen}
\rho_{DM} (t)=\frac{M^2 \cdot \chi^2_*}{2} \cdot \left(\frac{a_*}{a (t)} \right)^3 \simeq \frac{\left(\kappa M^{5} H_*\right)^{2/3}}{4\lambda} \cdot \left(\frac{a_*}{a(t)} \right)^3 \; ,
\end{equation}
where we used Eq.~(\ref{chistar}) in the last equality. We assume that the field $\chi$ constitutes all DM in the Universe. Hence, the abundance constraint at the matter-radiation equality must be satisfied:
\begin{equation}
\label{abundance}
\rho_{DM} (t_{eq}) \simeq \rho_{rad} (t_{eq}) \simeq \frac{\pi^2 g_{*} (T_{eq})}{30} \cdot T^4_{eq} \; ,
\end{equation}
where $g_* (T)$ is the effective number of ultra-relativistic degrees of freedom at the temperature $T$, and subscripts $'eq'$ and $'rad'$ stand for 
matter-radiation equality and radiation, respectively. 
Using Eqs.~(\ref{dmenergygen}) and (\ref{abundance}) one
can find the relation between the mass $M$ and the temperature $T_*$ of
the Universe at the moment $t_*$ (we assume the symmetry restoration at the radiation dominated stage):
\begin{equation}
\label{constraintgeneric}
M \simeq  \left(\frac{\lambda^3 }{75 \kappa^2} \right)^{1/10} \cdot \left(\frac{g_* (T_{eq})}{h_* (T_{eq})} \right)^{3/10} \cdot \left(4\pi^2 \cdot g_{*} (T_{*}) \right)^{1/5} \cdot T^{3/10}_{eq} \cdot T^{1/2}_{*} \cdot M^{1/5}_{Pl} \; ,
\end{equation}
where the reduced Planck mass is given by $M_{Pl}=2.43 \cdot 10^{18}~\mbox{GeV}$; $h_* (T)$ is the effective number of ultra-relativistic degrees of freedom entering the entropy density $s=2\pi^2 \cdot h_{*} (T) \cdot ~T^3/45$. 
To eliminate the Hubble rate $H_{*}$ from Eq.~(\ref{constraintgeneric}) we used Friedmann equation at the time $t_*$ assuming that DM component is subdominant,
\begin{equation}
\nonumber 
3M_{Pl}^2 H_*^2  = \frac{\pi^2 g_{*} (T_{*})}{30} \cdot T^4_{*} \; .
\end{equation}
Finally, to obtain Eq.~(\ref{constraintgeneric}) we also substituted the relation 
\begin{equation}
\label{ratioa}
\left(\frac{a_*}{a_{eq}} \right)^3 =\frac{h_* (T_{eq}) \cdot T^3_{eq}}{ h_{*} (T_{*}) \cdot T^3_*} \; .
\end{equation}
This expression follows from the entropy density conservation in the comoving volume: $s \cdot ~a^3 ~=\mbox{const}$. Note that $h_* (T)$ and $g_* (T)$ are slightly different only at the temperatures below the neutrinos decoupling, $T \lesssim 1~\mbox{MeV}$. In particular, we have
\begin{equation}
\label{hglow}
h_* (T_{eq}) \approx 3.9\;, \qquad g_* (T_{eq}) \approx 3.4 \; ,
\end{equation}
while at high temperatures,
\begin{equation}
\label{hghigh}
h_{*} (T \gg T_{SM}) =g_* (T \gg T_{SM}) \approx 106.75 \; ,
\end{equation}
so that $h_*(T_*) = g_*(T_*)$; here $T_{SM}$ is the temperature of the electroweak phase transition. 
We assume no ultra-relativistic degrees of freedom on top of the
Standard Model for times after the creation of DM, $t\geq t_*$.

Let us briefly mention constraints on the parameter space of the model of interest. The lower bound on the mass $M$ for given $\lambda$ follows from the 
limit on DM self-interaction~\cite{Markevitch:2003at, Tulin:2017ara}:
$\sigma/M \lesssim 1\,\text{cm}^2\text{/g} \approx 4.6 \cdot
10^{3}~\mbox{GeV}^{-3}$. Using $\sigma =9\lambda^2/(32 \pi M^2)$, one obtains $M \gtrsim \lambda^{2/3} \cdot
30~\mbox{MeV}$. From Eq.~\eqref{constraintgeneric}, where we use Eq.~\eqref{hglow} and $T_{eq} \approx 0.8~\mbox{eV}$, one finds that the mass~$M \simeq \lambda^{2/3}~\cdot 30~\mbox{MeV}$ 
corresponds to the temperature $T_* \simeq \lambda^{11/15}~\mbox{keV}$. There are reasons to expect that these lower bounds on the temperature $T_*$ and, consequently, $M$ can be further improved. 
Indeed, for the temperatures $T_* \lesssim 1~\mbox{keV}$, short wavelength perturbations responsible for the formation of relatively small structures in the Universe, may not have enough time 
to experience logarithmic growth during radiation domination. That is, by the comparison with a well established bottom-up picture of structure formation, one may conclude that $T_* \gg \lambda^{11/15} \cdot \mbox{keV}$. Keeping this in mind, we note 
that the simplest choice of the function $F$ considered in the next Section leads to large masses $M$ and temperatures $T_*$, so that there is no conflict with observational data.

An important qualification is in order here. When deriving Eq.~\eqref{constraintgeneric} we have assumed that DM is produced during radiation domination. 
This may not be the case for relatively low reheating temperatures and/or large masses $M$. 
DM may be created at even earlier stages, i.e., at preheating or inflation. 
In that case, Eq.~\eqref{constraintgeneric} is not applicable. We will study these scenarios and find the relevant expression in what follows.

So far we have focused on classical production of DM. However, there is also quantum-mechanical creation of DM particles around the time $t=t_*$, when adiabaticity condition is violated. 
Nevertheless, it can be shown that they give
a sub-dominant contribution to the total energy of DM. Indeed, the concentration of particles with conformal momenta in the range $(k, k+dk)$ produced quantum-mechanically is expressed via the Bogolyubov coefficient $\beta_k$: 
\begin{equation}
\nonumber 
dn_{\chi}=\frac{k^2 dk}{2\pi^2} \cdot \frac{1}{a^3(t)} \cdot |\beta_k|^2 \; .
\end{equation}
In the regime 
\begin{equation}
\label{ad}
\frac{\omega'_k}{\omega^2_k} \ll 1 \; ,
\end{equation}
where $\omega_k \approx \sqrt{k^2+a^2 M^2_{eff}}$ is the frequency of the mode with conformal momentum $k$, the Bogolyubov coefficient is given by~\cite{Mukhanov}
\begin{equation}
\label{bogolyubovad} 
\beta_k \approx \int^{\eta}_{\eta_i} d\tilde\eta \frac{\omega'_k}{2\omega_k} \cdot \mbox{exp} \left[-2i  \int^{\tilde\eta}_{\eta_i} d\tilde{\tilde{\eta}} \omega_k \right] \; .
\end{equation}
Here $\eta$ is the conformal time, $d\eta\equiv dt/a$, and $\eta_i$ corresponds to some point in the remote past with no DM particles; the prime $'$ denotes the derivative with respect to the conformal time. 
For modes with $k/a \gg M_{eff}$, the condition~\eqref{ad} is always satisfied, because 
$\omega_k \gg aM_{eff}$ in that case:
\begin{equation}
\nonumber 
\left|\frac{\omega'_k}{\omega^2_k} \right| \simeq \frac{a^3 M^3_{eff}}{\omega^3_k} \cdot \left|\frac{\dot{M}_{eff}}{M^2_{eff}} \right| \ll 1 \; .
\end{equation}
Therefore, for $k \gg a \cdot M_{eff}$, the expression~\eqref{bogolyubovad} is valid. Furthermore, for these large momenta the pre-exponential factor in Eq.~\eqref{bogolyubovad} changes
slowly on the time scale $\omega^{-1}_k$. Hence, the Bogolyubov coefficient is exponentially suppressed. 

Therefore, the largest momentum to be taken into account, when evaluating concentration of produced DM particles, is bounded as $k_{max}/a(t_*) \sim M_{eff} (t_*)$. 
For smaller momenta $k$ there may not be an exponential suppression. We can assume, however, that the Bogolyubov coefficient does not exceed unity by far for these momenta, i.e., $|\beta_k| \lesssim 1$. 
This assumption is supported by the comparison with particle creation in the regime of broad parametric resonance after inflation, where indeed $|\beta_k| \lesssim 1$. A scalar field, which plays the role of our field $\chi$ is produced non-perturbatively 
due to the interaction with an oscillating inflaton condensate. This interaction leads to oscillations of the effective mass of the scalar similar to those shown in Fig.~\ref{effectivemass}. We conclude that the energy density of DM particles is bounded as
\begin{equation}
\nonumber 
\rho_{DM, qm} \cdot \left(\frac{a(t)}{a_*} \right)^3 \lesssim \frac{M \cdot M^3_{eff} (t_*)}{6\pi^2} \simeq \frac{\kappa \cdot M^3 \cdot H_*}{6\pi^2} \; .
\end{equation}
Note the suppressing factor $2\lambda \kappa^{1/3} H^{1/3}_*/(3\pi^2 M^{1/3}) \ll 1$ relative to Eq.~(\ref{dmenergygen}). Hence, the energy density of quantum-mechanically produced particles gives 
a sub-dominant contribution to the total DM energy density.

\section{Non-minimal coupling to gravity}

In this Section we specify a concrete form of the function $F(x^\mu)$ resulting in the symmetry breaking at early times.
Perhaps, the simplest possibility is to choose $F$ to be proportional to the Ricci scalar $R$,
\begin{equation}
\label{concrete}
F(t)=-\frac{\xi}{\lambda} \cdot R \; ,
\end{equation}
where $\xi$ is some dimensionless coupling and $\lambda$ in the denominator is for the future convenience.
The cosmological value of the Ricci scalar can be expressed through the energy density of the dominant 
matter in the Universe $\rho$ and its equation of state $w$, using the trace of the Einstein equations, $M_{Pl}^2 R+T=0$, 
\begin{equation}
\nonumber 
R = -(1-3w) \cdot \frac{\rho}{M^2_{Pl}} \; .
\end{equation}
This relation can be rewritten in terms of the Hubble rate $H$,
\begin{equation}
\label{RH2}
R=-3\cdot (1-3w) \cdot H^2 \; ,
\end{equation}
where we used the Friedman equation $\rho=3H^2 M^2_{Pl}$. For the choice~\eqref{concrete} of the function $F(t)$, the parameter $\kappa$ defined from Eq.~\eqref{Frate} is given by 
\begin{equation}
\label{kappa}
\kappa = -\frac{2 \dot{H}}{H^2}=3(1+w)\,.
\end{equation}

Note that the choice~\eqref{concrete} implies the presence of $R^2$ term in the action~\eqref{modelbasic}. 
As a consequence a new degree of freedom, scalaron, appears.
The decay of the latter is capable of producing DM with the right
abundance~\cite{Gorbunov:2012ij, Gorbunov:2010bn, Arbuzova:2020etv}. 
However, in the present work we focus on a different mechanism of DM production 
resulting from the coupling to the Ricci scalar, therefore we would like to avoid the extra degree of freedom. 
To fulfill this goal, we slightly modify the model by eliminating the quadratic term in $F(x^\mu)$ from the action~\eqref{modelbasic}. With the use of Eq.~\eqref{concrete}, the modified action reads
\begin{equation}
\label{actionricci}
{\cal L}=\frac{(\partial_{\mu} \chi)^2}{2}-\frac{M^2 \chi^2}{2} -\frac{\lambda \chi^4}{4} -\frac{\xi}{2} \cdot \chi^2  \cdot R \; .
\end{equation}
Note that the analysis of Section~2 is applicable to the action~(\ref{actionricci}), since the form of the scalar field equation did not change. Namely, the field $\chi$ gets offset from its late minimum at $\chi=0$ due to the interaction with the 
Ricci scalar. As the Ricci scalar drops considerably, i.e., the following equality is obeyed:
\begin{equation}
\label{constraintonericci}
M^2 \simeq -\xi \cdot R_{*} \; ,
\end{equation}
the field $\chi$ starts to oscillate, and since then it acts as the standard DM. 
Depending on the moment of time, when 
this equality is reached, DM can be produced at the radiation-dominated stage, preheating or inflation.

We assume that the parameter $\xi$ is large, $\xi \gg 1$, so that the condition~\eqref{noisocurvature} is fulfilled.
The upper bound on the parameter $\xi$ is inferred from the assumption that the field $\chi$ remains sub-dominant during inflation. 
Clearly, the upper bound on $\xi$ depends on the choice of inflationary scenario. 
As an illustrative example let us assume Starobinsky inflation~\cite{Starobinsky:1980te}. The term $\xi \chi^2 R/2$ of our model should be small relative to the term $M^2_{Pl} R^2/6\mu^2$ of Starobinsky inflation, 
where $\mu \simeq 1.3 \cdot 10^{-5}~M_{Pl}$. Substituting the expectation value $\chi^2 =-\xi \cdot R/\lambda$, we obtain
\begin{equation}
\label{upperboundxi}
\xi \lesssim 3 \cdot \sqrt{\lambda} \cdot 10^{4} \; .
\end{equation}
Hence, in the weak coupling regime, $\lambda \lesssim 1$, the parameter $\xi$ is limited to be less than $3 \cdot 10^4$. 
A similar estimate~(\ref{upperboundxi}) is obtained in models of inflation driven by a scalar field, by requiring that the last term in Eq.~(\ref{actionricci}) is subdominant with respect to the Einstein-Hilbert term. In particular for $H_{infl}\sim 10^{13}$ GeV, we find that Eq.~(\ref{upperboundxi}) must be satisfied. 

Before we continue, it is worth to comment on studies in the literature involving the action of the form~\eqref{actionricci}. 
First, the non-minimal coupling 
to the Ricci scalar is common in inflation, with Higgs inflation~\cite{Bezrukov:2007ep} being the most notable example. In the context of DM, non-minimal coupling of DM field to the Ricci scalar 
has been discussed in Refs.~\cite{AlonsoAlvarez:2019cgw, Alonso-Alvarez:2018tus}. These references also deal with a version of misalignment mechanism for DM. However, the misalignment there is due to the slow roll 
of light DM field during inflation, rather than due to the interaction with the Ricci scalar. Finally, we note that the action of the exact form~\eqref{actionricci} has been considered 
in Ref.~\cite{Dimopoulos:2018wfg}, where the field analogous to $\chi$ is used to reheat the Universe.

\subsection{Dark Matter production during inflation}

If the masses $M$ are very large, up to the Grand Unification scale,
DM still can be produced abundantly in our model. 
Below we entertain this possibility.
Following the reasoning of Section~2, and using~(\ref{chistar}) and (\ref{kappa}), we find the amplitude
of the field $\chi$ at the onset of oscillations,
\begin{equation}
\chi_* \simeq \frac{\left(2 \epsilon_* M^{2} H_{*}\right)^{1/3}}{\sqrt{2\lambda}} \; ,
\nonumber 
\end{equation}
where $\epsilon_* = -\dot{H_*}/H_*^2$ is the slow roll parameter at the time $t_*$. 
The Hubble rate $H_{*}$ at the onset of oscillations is related to the
mass parameter by Eq.~\eqref{osconset}:  
\begin{equation}
\nonumber 
M^2 \simeq -\xi R_{*} \simeq 12 H^2_{*} \xi \; .
\end{equation}
The presence of $\xi \gg 1$ guarantees that the mass $M$ is much larger than the Hubble rate $H_*$, so that the condition~\eqref{MHcondition} is fulfilled.
For $\lambda \simeq 1$ and $H_{*} \simeq 10^{13}~\mbox{GeV}$, the upper bound~\eqref{upperboundxi} translates into the limit:
\begin{equation}
\label{supersuperheavy}
M \lesssim 6 \cdot 10^{15}~\mbox{GeV} \; .
\end{equation} 
Notably, DM with such a huge mass can be produced independently of the reheating temperature in the post-inflationary Universe, details of transition 
from inflationary to post-inflationary stage, particularities of preheating etc. 
Furthermore, because of the large initial amplitude of DM oscillations, the
initial energy density of DM is high. That is, were DM produced at the end of inflation, 
it would quickly overclose the Universe. However, once it is produced during inflation, there is a time 
for it to get considerably diluted.

We again assume that the field $\chi (t)$ constitutes all DM in the Universe. Hence, the abundance constraint must be obeyed: 
\begin{equation}
\nonumber 
\frac{\left(\epsilon_* M^{5} H_*\right)^{2/3}}{2^{4/3}\lambda} \cdot \left(\frac{a_*}{a_{eq}} \right)^3 \simeq \frac{\pi^2}{30} \cdot g_{*} (T_{eq}) \cdot T^4_{eq} \; . 
\end{equation}
It is convenient to split the ratio of scale factors as follows: 
\begin{equation}
\nonumber 
\left(\frac{a_*}{a_{eq}} \right)^3=\left(\frac{a_*}{a_e} \right)^3 \cdot \left(\frac{a_e}{a_{reh}} \right)^3 \cdot \left(\frac{a_{reh}}{a_{eq}} \right)^3 \; .
\end{equation}
Subscripts $'e'$ and $'reh'$ stand for the end of inflation and reheating, respectively. The ratio $a_{reh}/a_{eq}$ follows from Eq.~\eqref{ratioa}, where one should replace the subscript $'*'$ by $'reh'$. 
We estimate the ratio $a_e/a_{reh}$ assuming evolution during preheating alike matter domination:  
\begin{equation}
\nonumber 
\left(\frac{a_e}{a_{reh}} \right)^3 \approx \frac{\pi^2 \cdot g_* (T_{reh}) \cdot T^4_{reh}}{90 \cdot H^2_e \cdot M^2_{Pl}} \; .
\end{equation}
Combining these factors together along with the abundance constraint above, we obtain for the ratio $a_*/a_e$:
\begin{equation}
\nonumber 
\frac{a_*}{a_e} \simeq (3 \cdot \xi)^{1/9} \cdot \left(\frac{\lambda \cdot g_* (T_{eq})}{ h_* (T_{eq})} \right)^{1/3} \cdot \left(\frac{H_e}{\epsilon^{1/3}_* \cdot H_*} \right)^{2/3} \cdot \left(\frac{M^2_{Pl}}{12 \cdot \xi^2 \cdot H^2_*} \right)^{1/3} 
\cdot \left(\frac{T_{eq}}{T_{reh}} \right)^{1/3} \; .
\end{equation}
To estimate this expression, we choose the parameters as follows: $\lambda \simeq 1$, $\xi \simeq 3 \cdot 10^{4}$, $H_e \simeq \epsilon^{1/3}_* H_*$, and $12\xi H^2_* \simeq M^2_{Pl}$. Then, for reheating temperatures in the range $T_{reh} \simeq 10^{9}\!-\!10^{15}$\,GeV, we obtain that the field $\chi$ should enter the conventional oscillating regime very early, at the times (in terms of inflationary e-folds):
\begin{equation}
\nonumber 
N_* \simeq 16-20 \; .
\end{equation}  
To the best of our knowledge, this is the first model, which predicts DM creation at such early times.

Note that instead of non-minimal coupling of DM field to the the Ricci scalar, we can consider an interaction with the inflaton field $\varphi$: 
\begin{equation}
\label{intinflaton} 
{\cal L}=  \frac{\left(\partial_{\mu} \chi \right)^2}{2}-\frac{M^2 \chi^2}{2}-\frac{\lambda}{4} \cdot \left(\chi^2-\frac{g^2}{\lambda} \varphi^2 \right)^2 \; .
\end{equation}
For $g \lesssim 10^{-3}$, which is necessary to avoid generation of large loop corrections 
to the inflaton potential, one obtains a similar upper bound~\eqref{supersuperheavy} on the mass of DM produced during inflation. 
Note that the model~\eqref{intinflaton} has been considered in Ref.~\cite{Greene:1997ge}, which, however, focused on production of $\chi$-particles during (p)reheating. In that case, abundance of created $\chi$-particles largely exceeds that of DM. We do not consider this option in the present work.

\subsection{Dark Matter production during preheating}

Lighter DM, with the masses below the inflationary Hubble rate, can be produced efficiently during (p)reheating and at the radiation-dominated stage. 
In this Subsection we investigate the former option and we postpone a discussion of the latter to the next Subsection.
We assume that during preheating the equation of state is dust-like, $w=0$. Then, the Ricci scalar is given by 
\begin{equation}
\nonumber 
R=-12 H^2 -6\dot{H}=-3H^2 \; .
\end{equation}
Here we used the average values for the Hubble parameter, i.e., $H=2/3t$, and its derivative. 
In fact, at least at the beginning of preheating, both the Hubble parameter and the Ricci scalar undergo oscillations around its mean value, reflecting  oscillations of an inflaton. 
We discuss effects of these oscillations at the end of this Subsection. 

With this important qualification, our generic results of Section~2
are applicable. DM oscillations start at the time
$t_*$ defined from $M^2 \simeq -\xi R_{*}$. Upon substituting $R=-3H^2$ this relation translates into 
\begin{equation}
\label{linkpreheating}
M^2 \simeq 3\xi H^2_{*}\,. 
\end{equation}
The energy density of DM is given by
Eq.~\eqref{dmenergygen}, where we should substitute $\kappa=3$ according to Eq.~\eqref{kappa}. Combining the DM abundance constraint~\eqref{abundance} with Eq.~\eqref{linkpreheating}, 
we find
\begin{equation}
\label{H*}
H_{*} \simeq \frac{2~M_{Pl}}{3^{2/3}~ \xi^{5/6}}  \cdot \sqrt{\frac{ \lambda \cdot g_{*} (T_{eq}) }{h_{*} (T_{eq}) }} \cdot \sqrt{\frac{T_{eq}}{T_{reh}}} \; .
\end{equation} 
To extract this expression from the abundance constraint, we used the identity
\begin{equation}
\nonumber 
\left(\frac{a_*}{a_{eq}} \right)^3=\left(\frac{a_*}{a_{reh}} \right)^3
\cdot \left(\frac{a_{reh}}{a_{eq}} \right)^3 \;. 
\end{equation}
The first multiplier on the r.h.s. of the above expression can be found from the Friedmann equation at the reheating time,
\begin{equation}
\nonumber 
\left(\frac{a_*}{a_{reh}} \right)^3 \approx \frac{\pi^2 \cdot g_* (T_{reh}) \cdot T^4_{reh}}{90 \cdot H^2_* \cdot M^2_{Pl}} \,,
\end{equation}
while to obtain the second multiplier one can use Eq.~\eqref{ratioa} with the replacement $'*' \to 'reh'$.
Inserting Eq.~\eqref{linkpreheating} into Eq.~\eqref{H*} we obtain for the DM mass:
\begin{equation}
\label{masspreheating0}
M \simeq \frac{2 M_{Pl}}{(3\xi^2)^{1/6}} \cdot \sqrt{\frac{\lambda \cdot g_{*} (T_{eq}) }{h_{*} (T_{eq}) }} \cdot \sqrt{\frac{T_{eq}}{T_{reh}}} \; .
\end{equation}
Note that $H_*$ should be larger than $H$ at the onset of the hot stage, but lower than 
its value at inflation. The Hubble rate during inflation cannot exceed $H_{infl} \simeq
10^{13}~\mbox{GeV}$, given the limits on the relic gravitational waves~\cite{Akrami:2018odb}. As a result we have
\begin{equation}
\nonumber 
\sqrt{\frac{\pi^2 g_* (T_{reh})}{90}} \cdot \frac{T^2_{reh}}{M_{Pl}} \lesssim H_{*} \lesssim H_{infl} \; .
\end{equation}
This still allows for a fairly broad range for the reheating temperature: 
\begin{equation}
\nonumber 
\frac{40 \cdot \lambda}{\xi^{5/3}} \cdot \mbox{GeV} \lesssim T_{reh} \lesssim \frac{\lambda^{1/5}}{\xi^{1/3}} \cdot 5\cdot 10^{12}~\mbox{GeV} \; .
\end{equation} 
Hence, if the reheating temperature is in this range, DM with the
mass~\eqref{masspreheating0} can be produced with the right abundance
during preheating. From Eq.~\eqref{masspreheating0}, we find the corresponding range for the DM masses:   
\begin{equation}
\nonumber 
5\cdot\frac{\lambda^{2/5}}{\xi^{1/6}}~10^7~\mbox{GeV} \lesssim M \lesssim
2\cdot  \sqrt{\xi}\cdot10^{13}~\mbox{GeV} \; .
\end{equation} 

As it has been mentioned above, during preheating the Ricci scalar
generically oscillates around its mean value reflecting inflaton oscillations at preheating. 
These oscillations lead to production of DM particles~\cite{Fairbairn:2018bsw, Velazquez:2019mpj}. 
It turns out that in the parameter space of interest, the mechanism of Refs.~\cite{Fairbairn:2018bsw, Velazquez:2019mpj} is more efficient than ours. 
Hence, we face a problem of overproduction of DM, at least in common inflationary scenarios. One possible way out is to assume that the Ricci scalar 
is stabilized at negative values due to a very fast (almost instant) production of non-relativistic matter immediately as inflation terminates. 
This option is very restrictive, however. It is more natural to assume that DM interacts with a fermion singlet $S$ 
(e.g., sterile neutrino), which subsequently decays into Standard Model species:
\begin{equation}
\nonumber 
{\cal L}_{int} = g \chi \bar{S} S \; ,
\end{equation} 
where $g$ is a dimensionless coupling constant. Note that during preheating the effective DM mass is estimated by $M_{eff} \simeq \sqrt{\xi} H_{preh}$, where $H_{preh}$ is the Hubble rate during preheating. Hence, if the mass $M_S$ of the fermion $S$ is in the range $M \lesssim M_S \lesssim \sqrt{\xi} H_{preh}$, there is 
a decay channel of DM into a couple of fermions $S$. On the other hand, at relatively late times, well after the Ricci scalar stabilizes but before the time $t_*$, the decay of $\chi$ into fermions stops, and the field $\chi$ becomes stable. 
Note that for this scenario to be realized the coupling constant $g$ and/or $\xi$ should be relatively large. Indeed, for the early-time DM decay to be efficient, 
one should have 
\begin{equation}
\nonumber 
\Gamma_{\chi \rightarrow S} \simeq \frac{g^2 M}{8\pi} \simeq \frac{g^2 \sqrt{\xi} H_{preh} }{8\pi} \gtrsim H_{preh} \; .
\end{equation}
Otherwise, DM will be diluted only due to the cosmic drag. The above inequality gives a stringent constraint on the model parameter space:
\begin{equation}
\nonumber 
\sqrt{\xi} \gtrsim \frac{8\pi}{g^2} \; .
\end{equation}
To put it another way, the decay of $\chi \to S$ is efficient for large $\xi \gtrsim 10^{3}$ 
and not very small coupling constants $g$.

\subsection{Dark Matter production during radiation domination}

Finally, in this Section we consider DM production during radiation-domination, when $w = 1/3$. 
In this case the classical value of the Ricci scalar is zero, see Eq.~(\ref{RH2}). 
However, due to the conformal anomaly, $R$ does not vanish exactly,
and in fact $1-3w$ can reach values of the order $0.01-0.1$ at very high temperatures~\cite{Kajantie:2002wa}.
We observe below that the DM field $\chi$ indeed starts oscillating, when the Universe is very hot.

For the case at hand, the condition~\eqref{adiabaticity} translates into 
\begin{equation}
\label{adiabaticityricci}
\xi \gg \frac{1}{(1-3w)} \; ,
\end{equation}
i.e., we should assume relatively large $\xi$. The condition~(\ref{adiabaticityricci}) also leads
to suppression of isocurvature perturbations during inflation, 
see Eq.~\eqref{noisocurvature}.

Using Eqs.~\eqref{constraintgeneric} with $\kappa=4$ (see Eq.~\eqref{kappa}) and~\eqref{constraintonericci} we find the temperature $T_{*}$ and the mass $M$ as the functions of model parameters, $\lambda$ and $\xi$. The temperature $T_{*}$ is given by 
\begin{equation}
\nonumber 
T_{*} \simeq \left(\frac{324}{\pi^6} \right)^{1/15}  \cdot \frac{(10\cdot \lambda)^{1/5}}{g^{1/5}_{*} (T_{*}) \cdot \xi^{1/3} \cdot (1-3w)^{1/3}} \cdot \left[\frac{g_* (T_{eq}) }{h_{*} (T_{eq}) } \right]^{1/5}  
\cdot T^{1/5}_{eq} \cdot M^{4/5}_{Pl}\; . 
\end{equation} 
Substituting $g_*$ and $h_*$ from Eqs.~\eqref{hglow} and~\eqref{hghigh}, and using $T_{eq} \simeq 0.8~\mbox{eV}$, we obtain 
\begin{equation}
\nonumber 
T_{*} \simeq \frac{4 \cdot \lambda^{1/5} \cdot 10^{12}~\mbox{GeV}}{ \xi^{1/3} \cdot (1-3w)^{1/3}} \; .
\end{equation}
Note that the temperature $T_{*}$ is only moderately sensitive to the parameters $\xi$, $1-3w$ and depends very mildly on the coupling constant $\lambda$. Given the adiabaticity condition~\eqref{adiabaticityricci}, we 
obtain the upper bound on the temperature $T_*$: 
\begin{equation}
\nonumber 
T_* \ll 4 \cdot \lambda^{1/5}  \cdot 10^{12}~\mbox{GeV} \; .
\end{equation}
For not extremely small $\lambda$ we obtain that $T_* \simeq 10^{12}~\mbox{GeV}$.
These temperatures are predicted in some well motivated inflationary models, e.g., in Higgs inflation~\cite{Bezrukov:2007ep, Bezrukov:2019ylq}.

The mass of DM is constrained to be
\begin{equation}
\nonumber 
M \simeq \left(\frac{96 \pi^6}{125} \right)^{1/30} \cdot  \frac{1}{\left[\xi (1-3w)\right]^{1/6}}\cdot \left[\frac{\lambda \cdot g_{*} (T_{eq})}{h_{*} (T_{eq})} \right]^{2/5} \cdot g^{1/10}_{*} (T_*) \cdot  T^{2/5}_{eq} \cdot M^{3/5}_{Pl}  \; .
\end{equation} 
Using Eqs.~\eqref{hglow} and~\eqref{hghigh}, and substituting $T_{eq} \simeq 0.8~\mbox{eV}$, we obtain 
\begin{equation}
\nonumber 
M \simeq \frac{5\cdot \lambda^{2/5} \cdot 10^{7}~\mbox{GeV}}{\xi^{1/6} \cdot (1-3w)^{1/6}} \; .
\end{equation}
Once again, we observe a very soft dependence on the model parameters. Using the constraint~\eqref{adiabaticityricci}, we obtain the upper bound on the DM mass $M$ produced by our mechanism 
during radiation-domination: 
\begin{equation}
\nonumber 
M \ll 5 \cdot \lambda^{2/5} \cdot 10^{7}~\mbox{GeV} \; .
\end{equation}
We see that for not very small coupling constant $\lambda$, the DM mass is in the range $M \simeq 10^{6}-10^{7}~\mbox{GeV}$.

\section{Discussions}

In the present work, we suggested a version of misalignment mechanism of scalar DM production. The offset of the DM scalar field $\chi$ from zero occurs due to its non-minimal interaction with gravity, i.e., the Ricci scalar. 
In our model, for high curvatures (early times), the symmetry is spontaneously broken and the scalar field follows the minimum of the spontaneously broken phase.
As curvature drops, the symmetry restores and the DM field $\chi$ starts oscillating around zero. 
The simplest realization of gravitational misalignment mechanism assumes the interaction $\propto \chi^2 R$. 
In this scenario, superheavy DM is produced with the mass in the range $10^{6}~\mbox{GeV} \lesssim M \lesssim 10^{16}~\mbox{GeV}$.
The upper bound corresponds to DM production during inflation, while the lower bound stands for DM generated at radiation domination. It is important that the upper bound 
is independent of the reheating temperature in the early Universe, details of preheating or transition from the inflationary stage to preheating. This is a conceptual difference of 
our mechanism from other mechanisms of superheavy DM generation.

One disadvantage of the considered scenario is a lack of signatures in observational/experimental data. The reason is $Z_2$-symmetry, which forbids DM decay in our picture.
 To equip the model with some non-trivial phenomenology, one can assume a slight breaking 
of $Z_2$-symmetry. This can be done without spoiling the main idea---DM interacting directly only with gravity. 
For this purpose, one can introduce the following interaction of the scalar with the curvature:
\begin{equation}
\nonumber 
{\cal L}_{breaking}=\mu \cdot \chi \cdot  R \; ,
\end{equation}
where $\mu$ is some parameter of the mass dimension. This interaction leads to the decay of DM with potentially interesting phenomenology. The resulting decay rate into Standard Model particles is 
suppresed as $\Gamma \propto 1/M^4_{Pl}$. However, for large $M$ discussed in the present work and/or not extremely small $\mu$, the resulting lifetime of DM can be comparable with the observational lower bound $\tau \simeq \Gamma^{-1} \gtrsim 10^{28}~\mbox{s}$ at 90\% CL~\cite{Aartsen:2018mxl}.

\section*{Acknowledgments} 

We are indebted to Alexander Vikman for useful comments and discussions. The work of E.B. is supported by the CNRS/RFBR Cooperation program for 2018-2020 n. 1985 ``Modified gravity and black holes: consistent models and experimental signatures''. The work of D.G. is supported by the Russian Foundation for Basic Research grant 18-52-15001-NCNIa. The work of S.~R. has been supported by the Czech Science Foundation--GA\v CR, project 20-16531Y.

\end{document}